\documentclass{article}

\usepackage{arxiv}

\usepackage[utf8]{inputenc} 
\usepackage[T1]{fontenc}    
\usepackage{hyperref}       
\usepackage{url}            
\usepackage{booktabs}       
\usepackage{amsfonts}       
\usepackage{nicefrac}       
\usepackage{microtype}      

\usepackage{graphicx}
\usepackage{subfigure}

\title{From Federated Learning to Federated Neural Architecture Search: A Survey}

\author{
  Hangyu Zhu \\
  Department of Computer Science\\
  University of Surrey\\
  Guildford, GU2 7XH, UK \\
  \texttt{hangyu.zhu@surrey.ac.uk} \\
   \And
 Haoyu Zhang \\
  College of Information Science and Technology\\
  Donghua University\\
  Shanghai, 201620, China \\
  \texttt{zhy920816@sina.cn} \\
  \And
  Yaochu Jin\thanks{Corresponding Author} \\
  Department of Computer Science\\
  University of Surrey\\
  Guildford, GU2 7XH, UK \\
  \texttt{yaochu.jin@surrey.ac.uk} \\
}

\begin{document}
\maketitle

\begin{abstract}
Federated learning is a recently proposed distributed machine learning paradigm for privacy preservation, which has found a wide range of applications where data privacy is of primary concern. Meanwhile, neural architecture search has become  very popular in deep learning for automatically tuning the architecture and hyperparameters of deep neural networks. While both federated learning and neural architecture search are faced with many open challenges, searching for optimized neural architectures in the federated learning framework is particularly demanding. This survey paper starts with a brief introduction to federated learning, including both horizontal, vertical, and hybrid federated learning. Then, neural architecture search approaches based on reinforcement learning, evolutionary algorithms and gradient-based are presented. This is followed by a description of federated neural architecture search that has recently been proposed, which is categorized into online and offline implementations, and single- and multi-objective search approaches. Finally, remaining open research questions are outlined and promising research topics are suggested.

\end{abstract}

\section{Introduction}
\label{intro}
Deep neural networks (DNNs) have made great success in the fields of image classification, natural language processing, autonomous driving systems, and many others. However, designing DNNs with high-quality architectures usually requires to manually try a large number of different hyperparameters, which is always a tedious task requiring broad expertise in both machine learning and the application area. Therefore, neural architecture search (NAS) has become increasingly popular in recent years \cite{elsken2018neural}, which aims to automatically search for good neural architectures.

Conventional centralized learning systems, however, requires that all training data generated on different devices be uploaded to a server or cloud for training a global model, which may give rise to serious privacy concerns. To address this concern, federated learning \cite{yang2019federated} has been proposed to protect user's data privacy by communicating the model parameters or other model information instead of the raw data between the server and local devices. Naturally, performing NAS in a federated learning environment becomes of particular importance, although it is still in its infant stage.

This survey aims to provide an overview of research work both on federated learning and neural architecture search, focusing, however, on the emerging area of federated neural architecture search. We categorize federated learning systems into offline and online approaches, where online federated neural architecture search is more challenging due to additional requirements on the performance of the networks during the search process and stronger constraints on the computational resources. In addition, we briefly discuss the differences between single- and multi-objective search neural architecture search methods to highlight different ways of handling multiple objectives in federated learning, such as accuracy, communication costs, model complexity and memory requirements on the local devices. Finally, we outline the main remaining challenges in federated neural architecture search.

\section{Federated Learning}
Federated learning \cite{mcmahan2017communication} distinguishes itself from distributed learning in three aspects. First, the main purpose of federated learning is to protect user's private information while distributed learning aims to accelerate training speed. Second, federated learning cannot determine the data distribution of any client devices. By contrast, distributed learning is able to arbitrarily allocate subsets of the whole learning data. Finally, federated learning faces a more challenging training environment as it may contain millions of unbalanced participating clients whose connections to the server are probably unstable. For example, edge devices like mobile phones are frequently offline.

Federated learning is often categorized based on the distribution characteristics of the data \cite{yang2019federated} which is originally used in distributed learning. Strictly speaking, federated learning does not have the concept of 'the whole dataset', therefore, it is hard to accurately describe the federated data distribution to some extent as defined in distribute learning. In the following, we will discuss the data distributions in federated learning in greater detail.

\subsection{Horizontal Federated Learning}
\begin{figure}[!t]
\begin{minipage}[t]{1\linewidth}
\centering
\subfigure[Instance distributed learning]{
\begin{minipage}[b]{0.46\textwidth}
\includegraphics[width=1\textwidth]{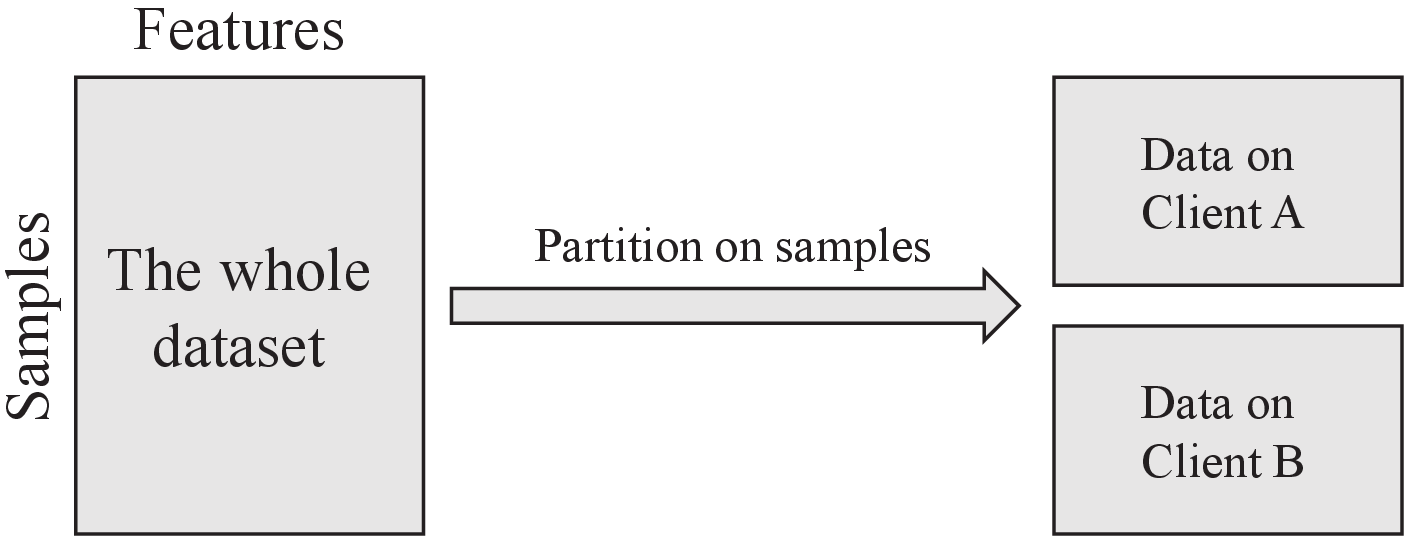}
\end{minipage}
}
\centering
\subfigure[Horizontal federated learning]{
\begin{minipage}[b]{0.46\textwidth}
\includegraphics[width=1\textwidth]{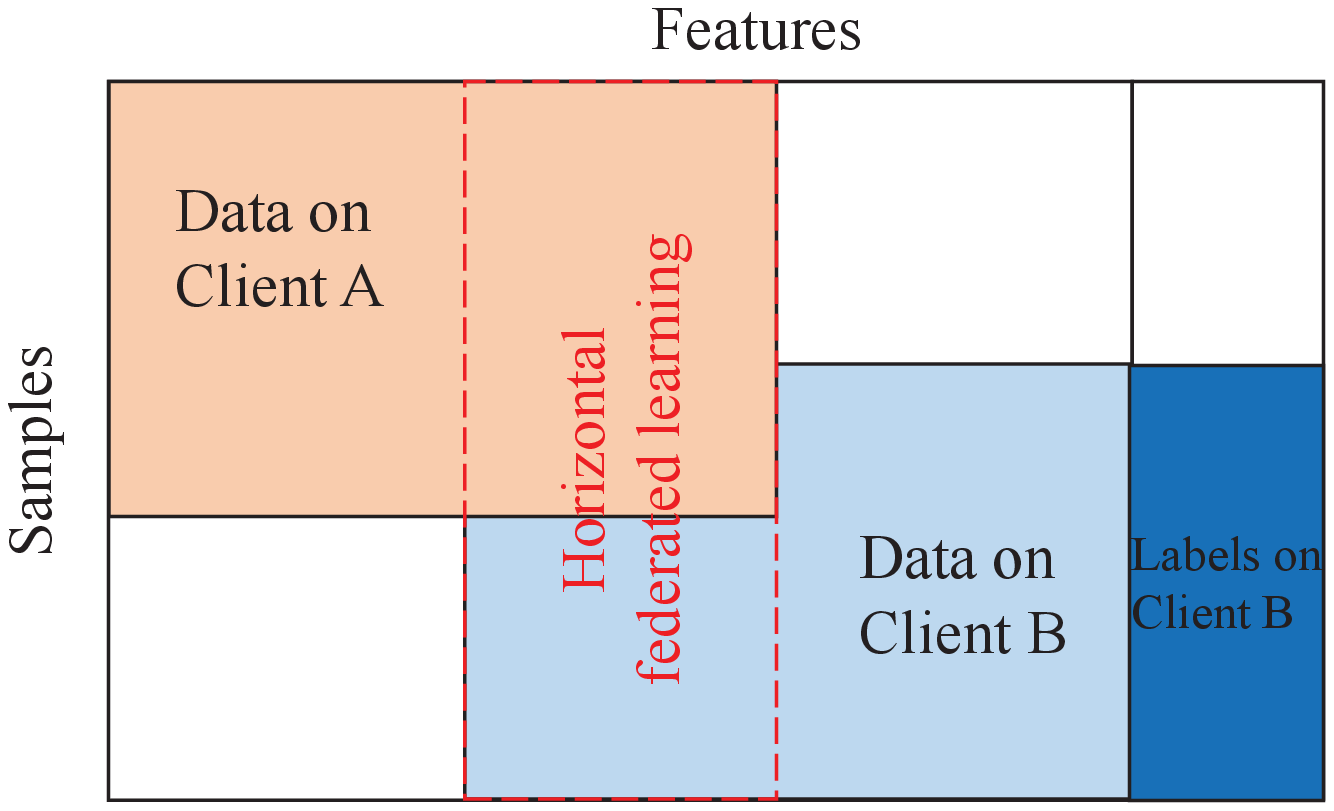}
\end{minipage}
} \\
\caption{Instance distributed learning (a) and horizontal federated learning (b).}
\label{horizontal}
\end{minipage}
\end{figure}

Horizontal federated learning is proposed for the scenarios in which datasets on the participating clients share the same feature space but have different samples. The name 'horizontal' originates from instance distributed learning \cite{zhang2018feature} as shown in Fig. \ref{horizontal}(a), where the whole dataset is horizontally partitioned over data samples and allocated to two clients. Similarly, as indicated by the part surrounded by the two dashed lines in Fig. \ref{horizontal}(b), the data can be considered as horizontally partitioned in federated learning, when different data are generated on different clients that have the same attributes (features). For instance, two hospitals in different regions may have different patients, although they performed the same tests for each patients and collected the same personal information such as the name, age, gender and address.

There are two main differences between instance distributed learning and horizontal federated learning. First, data is typically independent and identically distributed (IID) in distributed learning but may be non-IID in horizontal federated learning. As mentioned before, distributed learning is mainly designed for reducing the training time, therefore, designers can manually allocate every subsets of the client data to be IID to enhance the convergence. However, in horizontal federated learning, the central server has no access to any raw data, which are usually non-IID on different clients. Second, global model update mechanisms are different. Instance distributed learning, such as multi-GPU training a deep neural network, does not concern too much about the communication costs between the server and the clients. Once the gradients of mini-batch data on each client are calculated, they can be uploaded immediately to update the global model parameters on the server. This global model updating approach is intrinsically not suited for horizontal federated learning because frequent upload and download of data are not desirable due to the constraints on the communication costs.

\begin{figure}
\includegraphics[height=7cm, width=8cm]{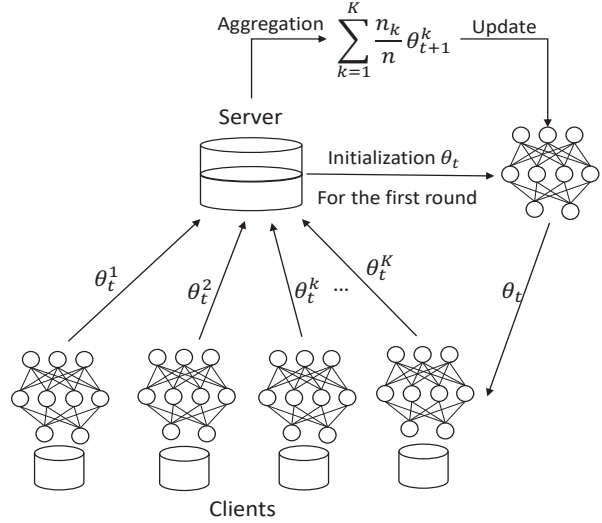}
\centering
\caption{Flowchart of federated learning. $ \theta  $ is the global model parameters, ${n_k}$ is the data size of client $k$, $K$ is the total number of clients and $t$ is the communication round in federated learning. We just initialize global model parameters randomly at the beginning of the communication round and use updated model parameters afterwards}
\label{hfederated}
\end{figure}

Typical horizontal federated learning (Fig. \ref{hfederated}) algorithms, such as the FedAvg proposed in \cite{mcmahan2017communication}, consist of the following main steps.
\begin{enumerate}
\item Initialize the global model parameters on the server and download the global model to every participating (connected) clients.
\item Every connected clients learn the downloaded global model on its own data for several training epochs. Once completed, the updated model parameters or gradients (gradients here means the difference between the downloaded model and updated model) are uploaded to the server.
\item The server aggregates all the upload the local models to update the global model.
\item Repeat the above two steps until convergence.
\end{enumerate}
From the above steps we can find that the central server can only receive model weights or gradients of the participating clients and has no access to any local raw data. Therefore, users' privacy is immensely protected in horizontal federated learning.

Horizontal federated learning has three additional main challenges compared to the traditional centralized learning: 1) it must reduce the communication resources as much as possible, 2) it needs to improve the convergence speed, and 3) it must make sure that no private information is leaked in passing the model information.  Much research work has focused on reducing communication costs, such as client updates sub-sampling \cite{shokri2015privacy,konevcny2016federated,caldas2018expanding} and model quantization \cite{DBLP:journals/corr/HanMD15,xu2020}. More recently, Chen \emph{et al.} \cite{chen2019communication} propose a layer-wise asynchronous update algorithm to reduce the communication costs by decreasing the update frequency of the deep layers in the neural network. In addition, Zhu \emph{et al.} \cite{zhu2019multi} use a multi-objective evolutionary algorithm (MOEA) to simultaneously enhance the model performance and communication efficiency. Learning a good model in horizontal federated learning is not an easy task, since the training data on different clients are usually non-IID, leading to possible model divergence. In order to solve this issue, Zhao \emph{et al.} \cite{zhao2018federated} empirically explore the effect of non-IID data and provide a statistical analysis of divergence. Li \emph{et al.} \cite{li2018federated} propose a FedProx algorithm to alleviate negative impacts of the system's heterogeneity by injecting a proximal term into the original loss on each client. Apart from it, an attentive aggregation method \cite{ji2019learning} is used to minimize the weighted distance between the server model and client models on non-IID datasets.

The central server is often regarded as honest but curious (follow federated learning protocol but try to infer client data information) in horizontal federated learning, and the revealed gradients of each client may potentially leak the data information \cite{shokri2015privacy}.
For this reason, Phong \emph{et al.} \cite{8241854} mathematically prove that model gradients (especially the first hidden weights) are proportional to the original data and adopt additive homomorphic encryption \cite{cryptoeprint:2015:1192} to encrypt and protect model gradients. In their method, the secret key is kept confidential to the server but known to all participating clients and the central server can easily get the plain model gradients as long as one of connected clients uploads its secret key. In order to mitigate this issue, secure multiparty computation (SMC) \cite{lindell2005secure,cryptoeprint:2017:281} is proposed to partition an intact secret key into several key shards and each client can just hold one shard. As a result, the server must get at least $t$ shards ($t$ is the threshold value) for decryption. Consequently, privacy preserving is significantly improved.

However, homomorphic encryption will increase the computation load, and SMC consumes much more communication resources, since encrypted model weights need to be downloaded and uploaded between the server and at least $t$ clients for partial decryption. Therefore, a more light-weight privacy preserving technique called differential privacy \cite{dwork2008differential} can also be used in horizontal federated learning. Such as the methods used in \cite{shokri2015privacy,geyer2017differentially}, a Gaussian or Laplacian noise is added to the gradients of each client before sending them to the central server. Note, however, that the learning process may be interrupted if the accountant \cite{abadi2016deep} exceeds a pre-defined threshold value. Most recently, Truex \emph{et al.} suggest a hybrid approach that combines differential privacy together with homomorphic encryption.

\subsection{Vertical Federated Learning}
\begin{figure}[!t]
\begin{minipage}[t]{1\linewidth}
\centering
\subfigure[Instance distributed learning]{
\begin{minipage}[b]{0.46\textwidth}
\includegraphics[width=1\textwidth]{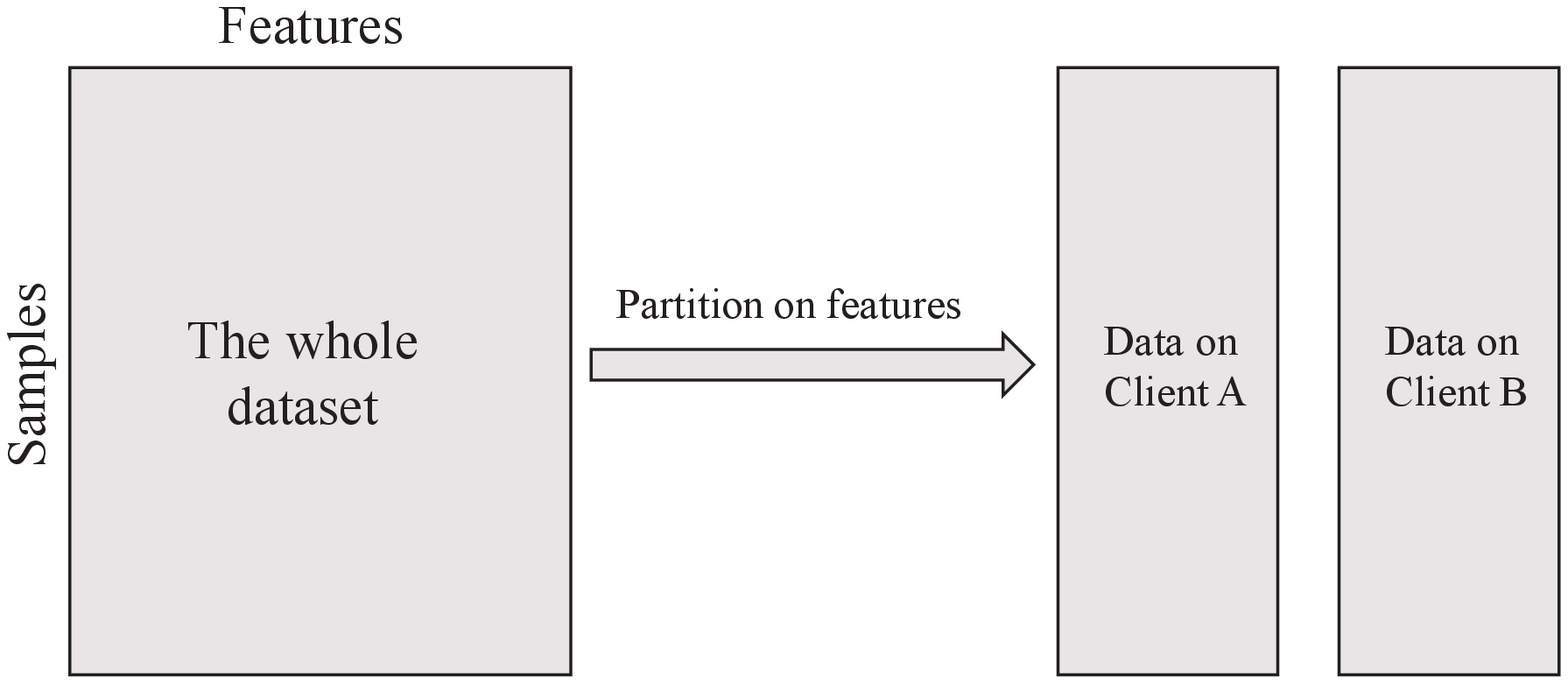}
\end{minipage}
}
\centering
\subfigure[Horizontal federated learning]{
\begin{minipage}[b]{0.46\textwidth}
\includegraphics[width=1\textwidth]{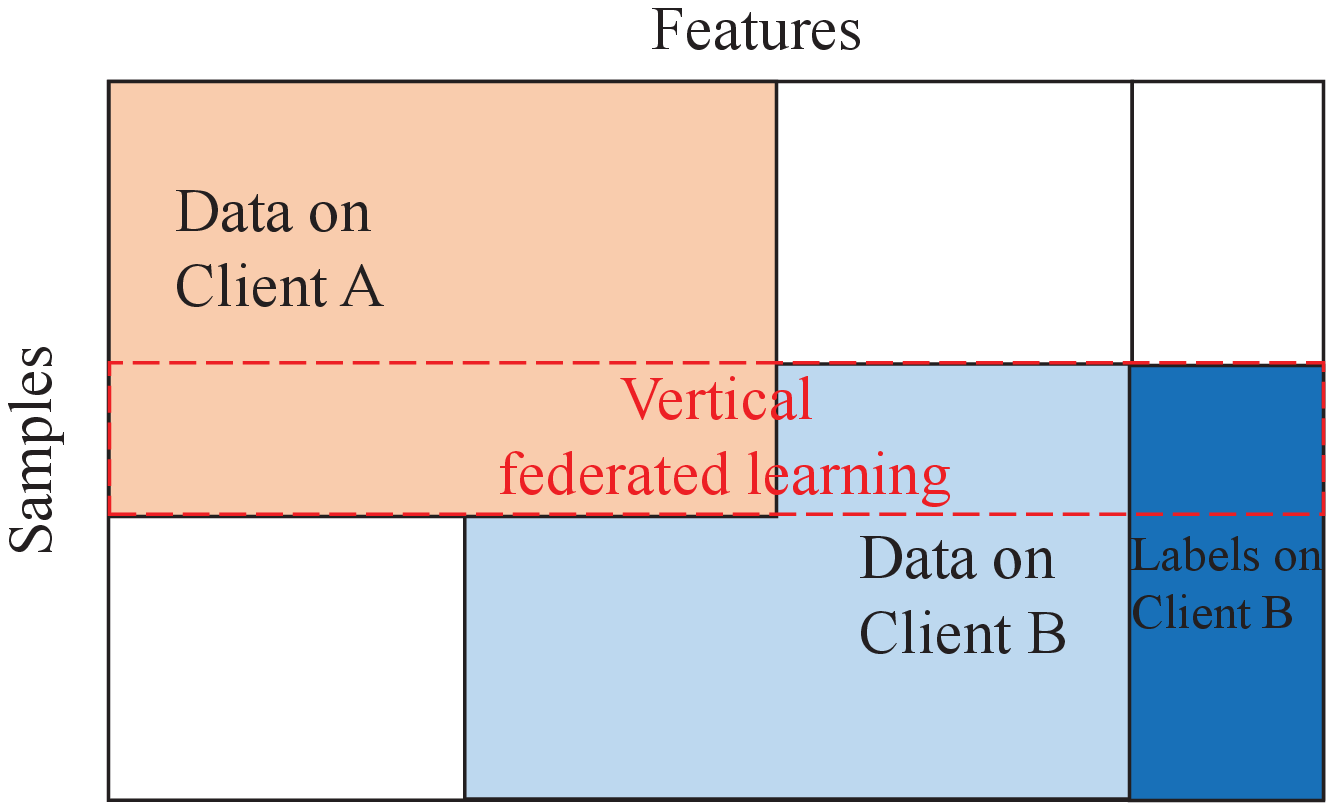}
\end{minipage}
} \\
\caption{Feature distributed learning (a) and vertical federated learning (b).}
\label{vertical}
\end{minipage}
\end{figure}

In contrast to horizontal federated learning, vertical federated learning is applicable to the situations where the datasets share the same sample space but have different feature space, as shown by part of surrounded by the dashed lines in Fig. \ref{vertical}(b). For example, two different financial agents may have the same customers but provide different services. Different from horizontal federated learning, vertical federated learning is similar to feature distributed learning \cite{zhang2018feature} to some extent which 'vertically' partitions the training data, as shown in Fig. \ref{vertical}(a) upon the feature space. Moreover, the central server is often called a coordinator \cite{hardy2017private} in feature distributed learning or vertical federated learning, since its main task is to calculate the total loss rather than aggregating the uploaded weights.

Vertical federated learning is firstly introduced in \cite{hardy2017private}, in which the overall framework contains one trusted coordinator and two parties, where each party represents one client. The coordinator computes the training loss and generates encryption key pairs. Homomorphic encryption is adopted for the privacy preserving purpose and the effect of entity resolution is also discussed. More recently, a two-party architecture \cite{yang2019parallel,liu2020secure} is proposed by removing the trusted coordinator which greatly reduces the complexity of the system. A typical two-party framework of vertical federated learning using a simple logistic regression model includes the following steps:
\begin{enumerate}
\item Assume Party A contains the data labels. Party A creates a homomorphic encryption key pair and sends the public key to Party B. Both parties initialize their local model parameters according to their feature dimensions of local training data.
\item Both parties compute their local inner products of data and the model. Then, Party B sends its results to Party A.
\item Party A sums two inner products and calculates the loss function by data labels. The loss is encrypted with a public key and is sent to Party B. The model gradients of Party A are also calculated.
\item Party B calculates the encrypted model gradients from the received loss and encrypt. In addition, a random number is encrypted and added to the encrypted gradients. The summation should be sent to Party A for decryption.

\item Party A uses a secret key to decrypt the summation value and sends it to Party B.
\item Update both model parameters of the two parties.
\item Repeat Step 2 to Step 6 until convergence.
\end{enumerate}

In Step 3, the training loss is encrypted before being sent to Party B, because it contains the information of the data labels in Party A which cannot be revealed to Party B. As a result, Party B needs to calculate its local model gradients on the encrypted loss and a Taylor approximation is commonly used \cite{hardy2017private,liu2020secure} for simplifying this computation.

From the above discussions, we can see that vertical federated learning is dramatically different from horizontal federated learning. The central server in horizontal federated learning is used for model aggregation, while in vertical federated learning the server plays the role of calculating the loss or collecting features. In addition, the server can be removed in vertical federated learning, e.g., summing the training loss within one of participating parties (clients). Apart from the above, we often assume that not all parties contain the data labels in vertical federated learning, e.g., only Client B contains data labels in Fig. \ref{vertical}(b) and those parties with no data labels are not able to update their models locally. Therefore, we call the server 'coordinator' that coordinates the feature predictions from all parties for calculating the training loss in vertical federated learning.

Most studies of vertical federated learning only support two parties (with or without a central coordinator) using a simple binary logistic regression model. Feng \emph{et al.} \cite{feng2020multi} adopt the idea of multi-view learning to extend the previous scheme into a multi-participant multi-class vertical federated learning framework. Besides, Liu \emph{et al.} introduce a federated stochastic block coordinate descent algorithm, where all participating parties update their local models for multiple times to reduce the total number of communication rounds. In addition, Chen \emph{et al.} propose an asynchronous vertical federated learning method and differential privacy is also used for privacy protection.

\subsection{Hybrid Federated Learning}
\begin{figure}[!t]
\begin{minipage}[t]{1\linewidth}
\centering
\subfigure[Symmetric federated learning]{
\begin{minipage}[b]{0.46\textwidth}
\includegraphics[width=1\textwidth]{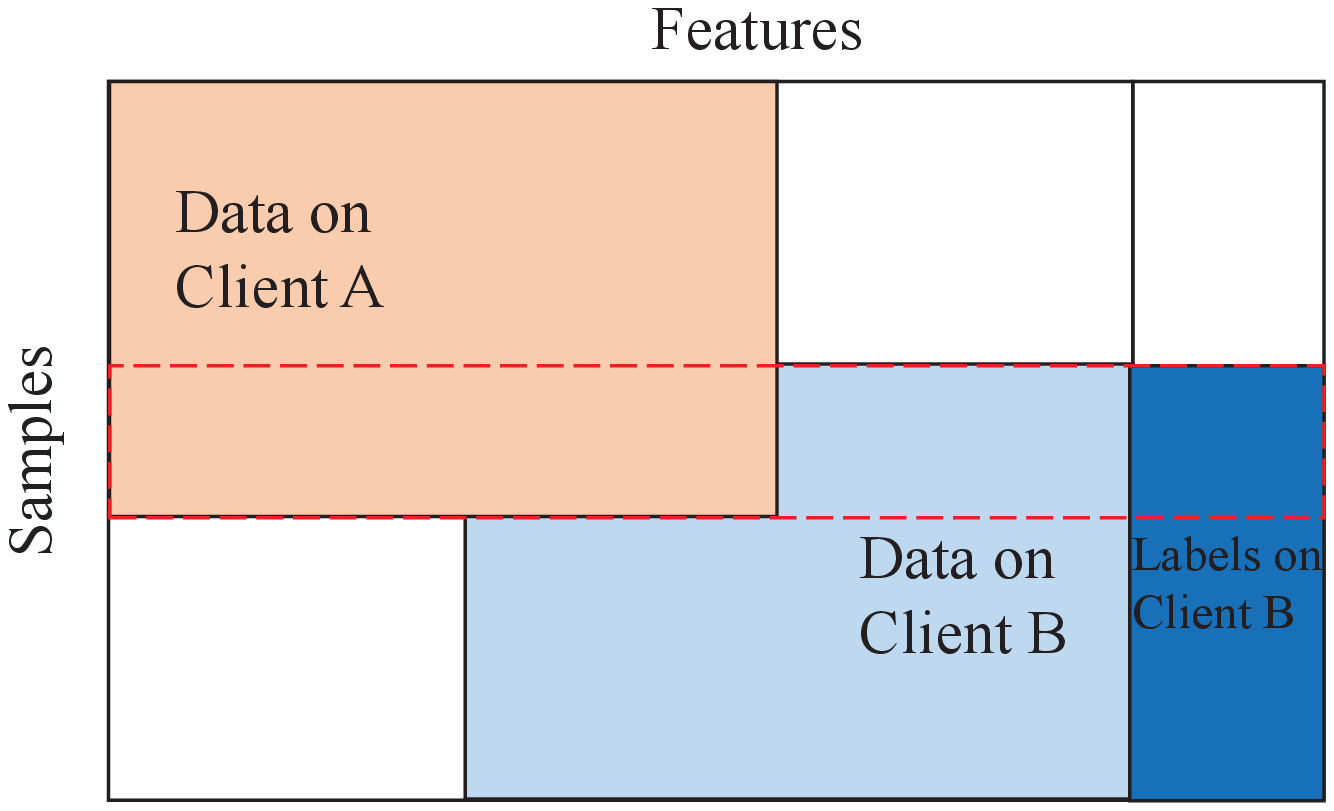}
\end{minipage}
}
\centering
\subfigure[Asymmetric federated learning]{
\begin{minipage}[b]{0.46\textwidth}
\includegraphics[width=1\textwidth]{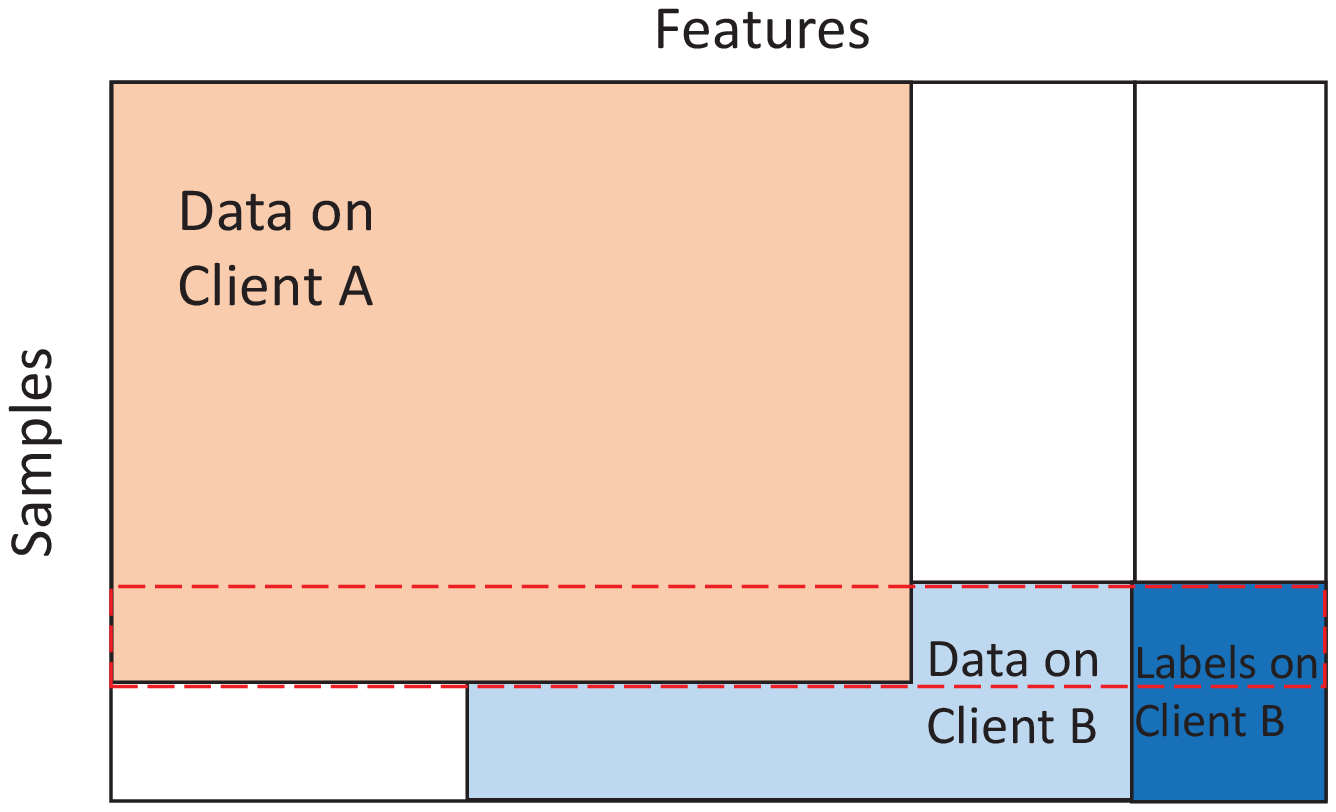}
\end{minipage}
} \\
\caption{a) Symmetric federated learning, and b) Asymmetric federated learning.}
\label{hybrid}
\end{minipage}
\end{figure}

Hybrid federated learning is more realistic in the real-world and it assumes that datasets on different clients not only have different sample spaces but also different feature spaces. Therefore, in this scenario, different parties need to share the data identity (ID) information to find the intersection part for distributed training, which is a threat to local clients' privacy. Since participants in hybrid federated learning are often asymmetric \cite{liu2020asymmetrically}, for instance, some participants are small companies always requiring to protect their ID information, while some participants are large companies that have no concern about the ID privacy. Symmetric federated learning and asymmetric federated learning are illustrated in Fig. \ref{hybrid}.

Secure ID alignment protocol is significant for hybrid federated learning, such as the commonly used Private Set Intersection (PSI) protocol. In standard PSI, all participants want to collaboratively find the intersection (the part indicated by the dashed lines in Fig. \ref{hybrid}) and keep unintersected parts private. The PSI protocols can be implemented by a classical public-key cryptosystem \cite{10.1007/978-3-540-24676-3_1,10.1007/978-3-540-25952-7_6} or other similar techniques.

Federated model training is similar to vertical federated learning, however, for asymmetric federated learning, Genuine with Dummy (GWD) approach \cite{liu2020asymmetrically} is used to ensure the correctness of computation results.

\section{Neural Architecture Search}

Since the quality of deep neural networks (DNNs) heavily depends on their architecture, increasing research efforts have been committed to design of novel structures in the deep learning community. However, manually designing deep neural networks requires considerable expertise in the field of deep learning and the investigated problem, which is unrealistic for many interested users. Not until recently has automated machine learning (Auto ML), in particular neural architecture search (NAS) become very popular to allow interested users without adequate domain knowledge to profit from the success of deep neural networks. The framework of NAS methods involves three dimensions \cite{elsken2018neural}, namely search space, search strategies, and performance estimation strategies.

The search space is a collection of network architectures, which has a major influence on the performance of the generated networks and search efficiency. The search strategy defines the method used to automatically design the optimal network architecture. To be specific, these search strategies can be divided into at least three categories reinforcement learning (RL), evolutionary algorithms (EAs), and gradient-based (GD) methods. In addition, a few additional methods, such as random search \cite{bergstra2012random,li2020random}, Bayesian optimization \cite{rasmussen2003gaussian,swersky2014freeze} and multinomial distribution learning \cite{zheng2019multinomial}, fall outside of the above categories. The search strategy aims at finding architectures that can obtain high performance on the test dataset. To guide searches effectively, these strategies utilize a performance estimation strategy to evaluate the quality of candidate architectures. Early work uses a simple way of performance estimation, for example, by iteratively training a candidate architecture on the  training dataset with the stochastic gradient descent (SGD) \cite{bottou2012stochastic} and evaluating its performance on the validation data \cite{zoph2016neural,zoph2016transfer,zoph2018learning,xie2017genetic,sun2019completely,sun2020automatically}. Such an evaluation strategy typically results in a prohibitively high computational cost. For example, to design a good performance of neural network, the automatic evolving convolutional neural network (AE-CNN) \cite{sun2019completely} algorithm consumes 27 GPU-days and the neural architecture search approach \cite{zoph2016neural} consumes 22400 GPU-days on the CIFAR10 dataset. Because inefficient search strategies require a large number of GPUs, many NAS methods cannot be implemented given limited computational resources. To address these challenges, much recent work dedicates to developing effective methods which can reduce the computational costs of performance evaluation, e.g., surrogate-assisted evolutionary algorithms (SAEAs) \cite{zela2018towards,swersky2014freeze,sun2019surrogate}, information reuse \cite{zhang2018finding,jin2018efficient}, one-shot neural architecture search \cite{pham2018efficient,guo2019single,dong2019one,li2020improving,you2020greedynas}, among many others.

\begin{figure}
\centering
\includegraphics[height=7.5cm, width=0.6\textwidth]{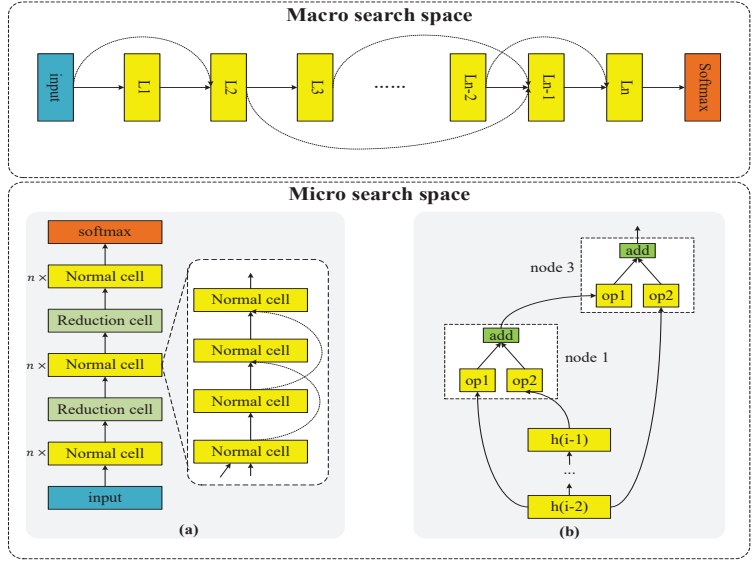}
\caption{Illustration of the marco and micro search spaces.}
\label{Fig_7}
\end{figure}

\subsection{NAS based on Reinforcement Learning}
Early work on NAS depends on RL to search for high-performance neural architectures \cite{zoph2016neural,zoph2016transfer,zoph2018learning}. The design of a network model is considered as an agent's action, which specifies the architecture of the network (i.e., a child model). The network is then trained and its performance on the validation data is returned as the agent's reward.

\begin{figure}
\centering
\includegraphics[width=0.92\textwidth]{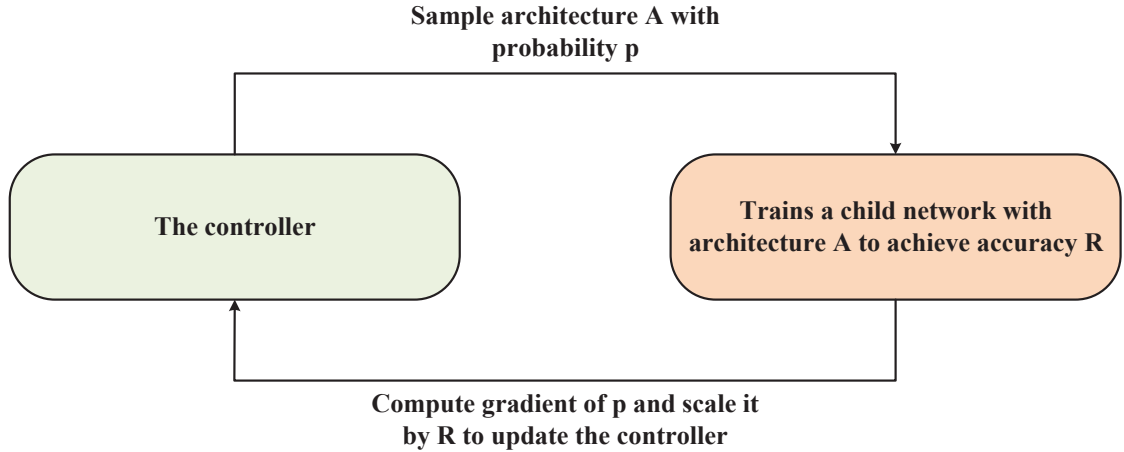}
\caption{An overview of RL-based NAS method.}
\label{Fig_4}
\end{figure}

A policy gradient method has attempted to approximate some nondifferentiable reward function to train a model which needs parameter gradients. Zoph \emph{et al.} \cite{zoph2016neural,zoph2018learning} first adopt this algorithm in NAS to train a recurrent neural network (RNN) model that generates architectures. As is shown in Fig. \ref{Fig_4}, the controller as a navigating tool to find more promising architectures in the search space. The original method in \cite{zoph2016neural} uses a macro search space that generates the entire network at once. As is shown in Fig. \ref{Fig_7}, the whole architecture consists of $n$ sequential layers where the dashed lines indicate skip connections. Hence, the macro search space aims to design the entire CNN architecture in terms of the number of hidden layers $n$, operations types (e.g. convolution), network hyper parameters (e.g., the convolutional kernel size), and the link methods (e.g. skip connections). However, this method is expensive when the data set is large. To reduce the computational cost, Zoph \emph{et al.} \cite{zoph2018learning} propose a more structured search space, called micro search space. The micro search space only covers repeated smaller modules, called normal cell and reduction cell, and then connects them together to form an entire network. As shown in Fig. \ref{Fig_7}, these cells are built in complex multi-branch operations (e.g. convolution). Each cell structure contains two inputs $h[i-1]$ and $h[i-2]$ coming from two previous layers. Hence, the micro search space aims to design structures of these two types of cells. In addition, the cell structures should have a good capability of generalizing to other related tasks. For example, the proposed method searches for optimal cell structures on the CIFAR10 data set and transfers them to the ImageNet data set by stacking together multiple copies of this cell. After that, the NASNet \cite{zoph2018learning} method is extended to a multi-objective optimization variant to simultaneously optimize the classification performance and computational cost using different scalarization parameters.

Q-learning \cite{watkins1989learning}, as a class of popular RL methods, has been widely used for NAS. Baker \emph{et al.} \cite{baker2016designing} employ an $\epsilon$-greedy Q-learning strategy to train a policy that sequentially chooses a type of layers (e.g. convolutional layer, pooling layer, and fully connected layer) and their corresponding hyperparameters. Zhong \emph{et al.} \cite{zhong2020blockqnn} extend this method to a block-wise network generation approach, which designs blocks with the same Q-learning paradigm. After that, the optimal blocks are repeated and stacked to construct the entire network architecture. To accelerate the search process, a distributed asynchronous strategy and an early-stop approach are adopted.

Parameter sharing introduced in efficient NAS (ENAS) \cite{pham2018efficient} is a promising approach for speeding up the search process for RL-based NAS methods, which treats architectures as different sub-graphs (sub-net) of a larger graph (super-net) and forces all sub-graphs to share a common set of weights that have edges of this larger graph in common. Pasunuru \emph{et al.} \cite{pasunuru2019continual} propose a multi-task architecture search (MAS) approach based on ENAS \cite{pham2018efficient} for finding a cell structure that performs well across multiple tasks. Hence, the cell structure generated by NAS can transfer to a new task. Bender \emph{et al.} \cite{bender2020can} propose a thorough comparison between random search NAS methods and ENAS \cite{pham2018efficient} on a larger search spaces for image detection and classification tasks, respectively. In addition, a new reward function is suggested, which can effectively improve the quality of the generated networks and reduce the difficulty for manual hyperparameter tuning. Liu \emph{et al.} \cite{liu2020search} present a novel knowledge distillation \cite{hinton2015distilling} approach to NAS, called architecture-aware knowledge distillation (AKD), which finds student models (compressed teacher models) that are best for distilling the given teacher model. The authors employ a RL-based NAS method with a KD-guided reward function to search for the best student model based on a given teacher model.

\subsection{NAS based on EAs}
\begin{figure}
\centering
\includegraphics[width=0.92\textwidth]{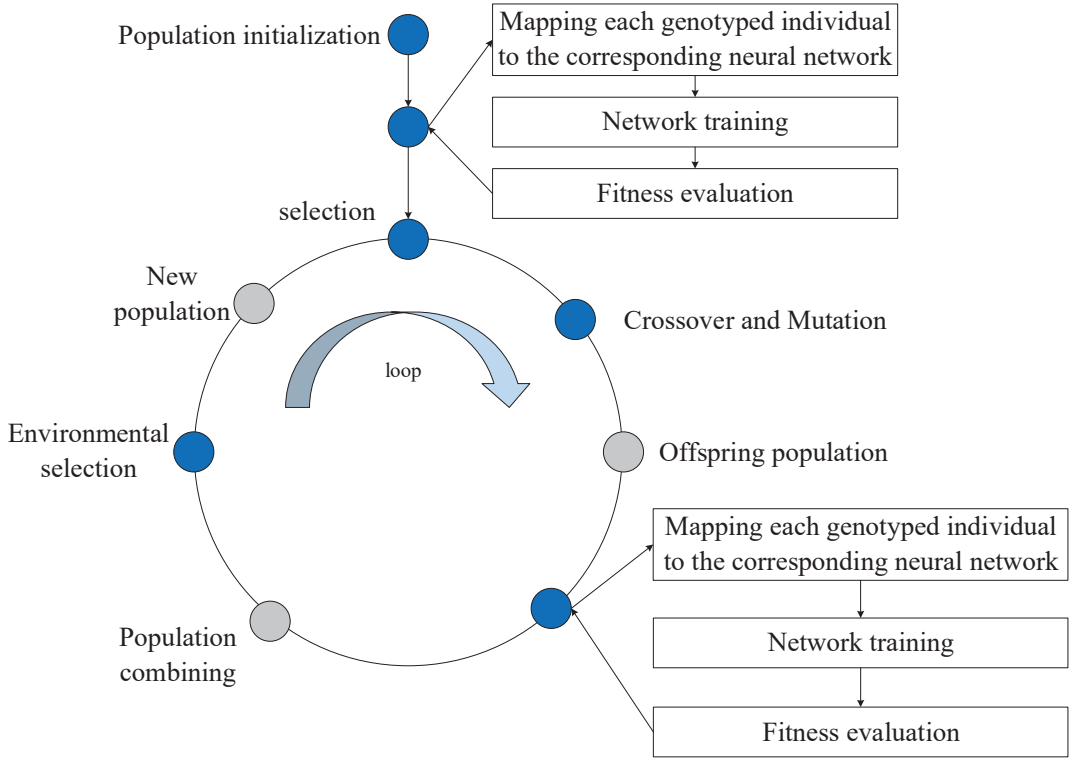}
\caption{A generic EA-based NAS framework.}
\label{Fig_5}
\end{figure}

EAs are a class of population-based, gradient-free heuristic search paradigms, which have been widely used in solving various complex optimization problems \cite{back1996evolutionary,banzhaf1998genetic,schmitt2001theory,zhu2019multi}. Historically, EAs have already been used for simultaneous optimization of the topology, weights of the connections and hyperparameters of artificial neural networks (ANNs) \cite{yao1999evolving,floreano2008neuroevolution,stanley2009hypercube,jozefowicz2015empirical,floreano2008neuroevolution}. The neuroevolution with augmenting typologies (NEAT) algorithm \cite{stanley2002evolving} is one of the popular early methods that have shown powerful performance. However, the traditional approaches are not well suited for optimizing DNNs due to the complex network architectures and large quantities of connection weights. EA-based NAS approaches to optimizing deep network architectures have started gaining momentum again recently \cite{galvan2020neuroevolution,darwish2020survey,stanley2019designing}, mainly because they can simultaneously explore multiple areas of the search space and their relative insensitiveness to a local minimum \cite{wang2014two_arch2,sun2018igd}. Fig. \ref{Fig_5} shows a generic framework of EA-based NAS algorithms. Broadly speaking, the whole process of an EA-based NAS algorithm follows the procedure of an EA containing at least four-steps: population initialization, offspring generation, fitness evaluation, and environmental selection. Generally, each neural network in the search space is encoded as a chromosome, and crossover and mutation operations of the chromosomes are performed in the exploration. Then, each chromosomes is transformed into a corresponding neural network, and iteratively trained on the training dataset. The trained network is evaluated on the validation dataset to get their fitness value.

Xie \emph{et al.} proposed a genetic CNN \cite{xie2017genetic} method that is one of the early studies using an EA for optimizing convolutional neural networks (CNNs). The genetic CNN algorithm searches over the entire architecture space and employs a fixed-length binary string to represent the connection between a number of ordered nodes (e.g. a convolutional operation). Although this early algorithm has some limitations, including a limited number of nodes as well as limited sizes and operations of convolutional filters, the generated structures have not only achieved competitive results on the CIFAR and SVHN datasets, but also shown excellent transferability to the ImageNet dataset \cite{russakovsky2015imagenet}.

Miikkulainen \emph{et al.} \cite{miikkulainen2019evolving} propose a coevolution DeepNEAT (CoDeepNEAT) method by extending the NEAT algorithm \cite{stanley2002evolving} to DNNs. In CoDeepNEAT, each neural network is assembled by modules and blueprints. A coevolutionary method is adopted that evolves two populations of modules and blueprints separately, in which each module chromosome represents a small DNN. The blueprints chromosome represents a graph where each node contains a pointer to a particular module species. The assembled networks are trained and evaluated in an ordinary way of NAS. The fitness of the network is the average fitness of the entire candidate models containing the blueprints or modules. In addition, Liang \emph{et al.}  found that the CoDeepNEAT also achieves promising performance in the Omniglot multi-task learning domain \cite{liang2018evolutionary}.

In fact, the length of a chromosome usually represents the depth of the corresponding neural network and a fixed encoding scheme may limit the performance of the optimized network. To address this issue, some recent NAS algorithms have attempted to use a variable-length encoding scheme. Real \emph{et al.} \cite{real2017large} propose a large-scale evolutionary NAS method, which utilizes a variable-length encoding method in which the network architectures can adaptively change their depths. Sun \emph{et al.} \cite{sun2019completely} propose an AE-CNN algorithm that can fully automatically design CNN architectures, without requiring any pre-processing or post-processing. Inspired by the ResNet \cite{he2016deep} and DenseNet \cite{huang2017densely}, AE-CNN's search space is defined by some predetermined building blocks, including ResNet block and DenseNet block, max pooling layer and mean pooling layer. Then, the authors design an EA-based NAS framework, including the variable-length encoding and a novel crossover and mutation operators based on the variable-length encoding, as the search strategy to search the optimal depth of the CNN. Given the nature of the variable-length encoding strategy, the algorithm employs a repair mechanism that avoids to produce invalid CNNs. Inspired by directed acyclic graph (DAG), William \emph{et al.} \cite{irwin2019graph} introduce a DAG-based encoding strategy, which can represent CNNs of an arbitrary connection structure and an unlimited depth.

Suganuma \emph{et al.} \cite{suganuma2017genetic} propose a CGP-CNN algorithm to design CNN architectures using genetic programming. The search space of CGP-CNN is represented by a DAG, where the nodes represent either convolutional blocks or concantenation operations. Then, CGP-CNN uses the Cartesian genetic programming (CGP) \cite{harding2008evolution,miller2006redundancy} encoding scheme to represent network architectures and their connectivity. This encoding scheme can represent variable-length network architectures and skip connections.

Most EA-based NAS methods aim at finding better topologies for DNNs while leaving the learning of weights to SGD. It is known that the SGD optimizer heavily rely on the initial values of the weights. To alleviate this problem, Sun \emph{et al.} \cite{sun2019evolving} propose an EA-based NAS method, named Evolving Deep CNNs (EvoCNN), to automatically design CNN architectures and corresponding connection weight initialization values without manual intervention. To reduce the search space, two statistical measures, including the mean and standard deviation of the connection weights, are encoded in the chromosome to represent tremendous numbers of the connection weights. In addition, the incomplete training scheme is employed to accelerate the fitness evaluation. According to the Occam’s razor theory \cite{rasmussen2001occam}, the number of connection weights are also considered as an indicator to scale the quality of candidate networks.

Sun \emph{et al.} \cite{sun2020automatically} use a genetic algorithm (GA) to design CNN architectures (CNN-GA). In CNN-GA, the standard convolutional layer is replaced by a novel building block, called the skip layer. The skip layer consists of two convolutional layers and one skip connection. Hence, the genotype encodes information of the skip layers and pooling layers. The fully connected layers are discarded, mainly because they easily result in the overfitting  \cite{hawkins2004problem}.

Rather than generating the entire CNNs, the micro search space \cite{pham2018efficient} has also been successfully employed by many recent EA-based NAS algorithms \cite{liu2018hierarchical,lu2019nsga,real2019regularized,zhou2020econas,lu2019multi}. Real \emph{et al.} \cite{real2019regularized} propose an extension  of the large-scale evolution \cite{real2017large}, called AmoebaNet, which has achieved better results on ImageNet compared with hand-designed methods for the first time.

Since EAs are a class of population-based search methods, the main computational bottleneck of EA-based NAS approaches lies in evaluating the fitness of the individuals by invoking the lower-level weight optimization. One such evaluation typically takes several hours to days if the network is large and if the training dataset is huge. For instance, on the CIFAR10 datasets, the AE-CNN \cite{sun2019completely} consumed 27 GPU days,  CNN-GA \cite{sun2020automatically} consumed 35 GPU days, and the large-scale evolutionary algorithm \cite{real2017large} consumed 2750 GPU days, AmoebaNet \cite{real2019regularized} consumed 3150 GPU days. This seriously limits the practical usability of most evolutionary NAS methods under a constrained search budget.

Therefore, various techniques have been suggested to accelerate the fitness evaluation, such as information reuse \cite{zhang2018finding,guo2019single} and SAEAs \cite{jin2011surrogate}. SAEAs have been popular to solve computationally expensive optimization problems, which use cheap classification and regression models, e.g., radial basis function networks (RBFNs) \cite{broomhead1988radial} and Gaussian process (GP) models \cite{jeong2005efficient}, to replace the expensive fitness evaluation \cite{wang2017committee}. Generally, the candidate networks are trained from a few number of expensive fitness evaluations, and then the trained networks are used to build a fitness predictors to reduce the cost of fitness evaluations. In the area of evolutionary NAS, Swersky \emph{et al.} \cite{swersky2014freeze} adopt Bayesian optimization \cite{shahriari2015taking} to speed up evolutionary optimization, which is called Freeze-thaw Bayesian optimization. Unfortunately, this algorithm is based on Markov chain Monte Carlo sampling and also suffers from high computational complexity. Sun \emph{et al.} proposed a performance predictor termed E2EPP, which is based on a class of SAEAs method \cite{sun2019surrogate} meant for offline data-driven evolutionary optimization of expensive engineering problems. Specifically, E2EPP builds a surrogate that can predict the quality of a candidate CNN, thereby avoiding the training of a large number of neural networks during the search process. Compared with AE-CNN, a variant of AE-CNN assisted by E2EPP (called AE-CNN+E2EPP) can reduce 230\% and 214\% GPU days on CIFAR100 and CIFAR10, respectively.

Knowledge inheritance \cite{zhang2018finding,guo2019single} is another promising approach to accelerate fitness evaluations. Zhang \emph{et al.} \cite{zhang2018finding} propose an EA based on asexual reproduction to find better typologies for deep CNNs and knowledge inheritance to reduce the computation cost. Once the topology of an offspring individual is generated by its parent, the weights of offspring networks are directly copied from its parents. For edges that do not appear in its parent network, the weights are randomly initialized.

To reduce the computational burden for fitness evaluations, another widely adopted approaches are to train and evaluate individuals using proxy metrics \cite{zhou2020econas,real2019regularized,lu2019multi}. The performance of the proxy models is used as the surrogate measurements to guide the evolutionary search. Such proxy metrics include reducing the width (the number of channels) and the depth (number of layers) for the intended network architecture to create a small-scale network, shortening the training time, reducing the resolution of input images, and training on a subset of the full training dataset. However, these simple proxy model constructing methods may result in a low correlation in prediction mainly because they may introduce biases in fitness estimation. Zhou \emph{et al.} \cite{zhou2020econas} have conducted extensive experiments on different combinations of proxy metrics to investigate their behaviors in maintaining the rank consistency in NAS, based on which a reliable hierarchical proxy strategy is proposed to accomplish economical neural architecture search (EcoNAS). The hierarchical proxy strategy aims at discarding less promising candidate individuals earlier with a fast proxy and estimates more promising individuals with a more expensive proxy. Hence, the EcoNAS method is able to achieve a $400 \times $ reduced search time in comparison to AmoebaNet \cite{real2019regularized} without sacrificing the performance. Lu \emph{et al.} \cite{lu2019multi} empirically establish the trade-off between the correlation of proxy performance to true performance and the speed-up in estimation.

Evolutionary multi-objective NAS methods considering multiple conflicting objectives have been reported. One of the earliest evolutionary multi-objective methods to design CNNs is NEMO \cite{kim2017nemo}, which simultaneously optimizes classification performance and inference time of a network based on NSGA-II \cite{deb2002fast}. Inspired by NEMO, Lu \emph{et al.} \cite{lu2019nsga} consider classification error and computational complexity as the two objectives. In addition, they empirically test multiple computational complexity metrics to measure the inference time of a network containing the number of active layers, the number of activating connections between layers, and the number of floating-point operations (FLOPs). And then, the FLOPs are used as a second conflicting objective for optimization. Moreover, a Bayesian network (BN) is adopted to learn the knowledge about promising architectures present in the search history and then guide the future exploitation in generating new architectures. Subsequently, Lu \emph{et al.} suggest NSGANet-v2 \cite{lu2019multi}, an extension of NSGANet \cite{lu2019nsga}, where a comprehensive search space including more layer operations and one more option that controls the width of the model is introduced. Dong \emph{et al.} \cite{dong2018dpp} present a DPP-Net on the basis of \cite{liu2018progressive} that optimizes both GPU memory usage and the model performance. Elsken \emph{et al.} \cite{elsken2018efficient} proposed the LEMONADE method, which formulates the NAS as a bi-objective optimization problem that maximizes the performance and minimizes the required computational resources. Inspired by \cite{wei2016network}, LEMONADE reduces computational cost through a custom-designed approximate network morphisms, which makes offspring individuals to share weights with their forerunners, avoiding training new networks from scratch. Note that evolutionary multi-objective structure optimization of shallow networks can be traced back to a decade ago \cite{jin2008pareto}.

\subsection{NAS based on GD}
\begin{figure}
\centering
\includegraphics[width=0.92\textwidth]{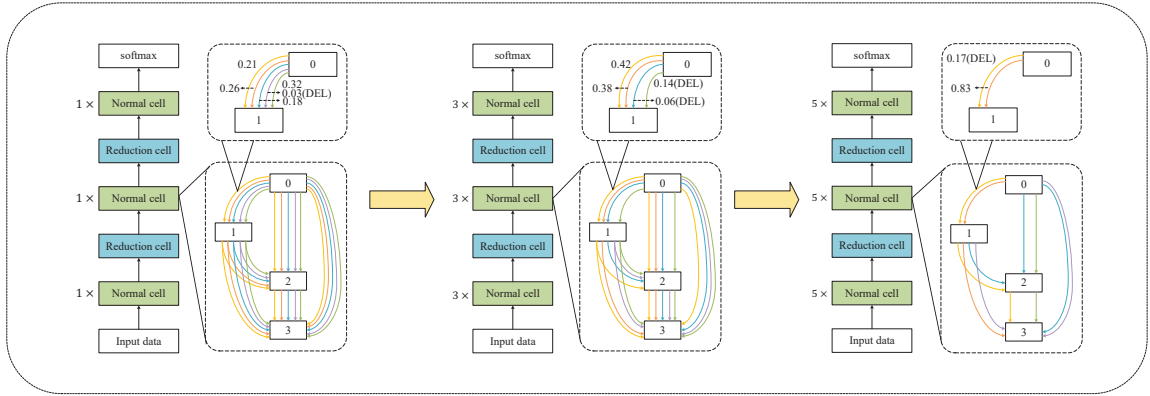}
\caption{A generic pipeline GD-based NAS method.}
\label{Fig_6}
\end{figure}

Compared with the above gradient-free optimization methods, the GD-based methods (Fig. \ref{Fig_6}) have become increasingly popular recently, mainly because their search speed is much faster than RL-based and EA-based methods. Early GD-based methods \cite{shin2018differentiable,ahmed2018maskconnect,fang2020densely,cai2018proxylessnas} implement this idea for optimizing layer hyperparameters or connectivity patterns, respectively. Lorraine \emph{et al.} \cite{lorraine2020optimizing} introduce an algorithm for inexpensive GD-based hyperparameter optimization. Liu \emph{et al.} \cite{liu2018darts} employ GD in the DARTS algorithm, which optimizes both the network weights and the architecture. The authors use relaxation tricks to make a weighted sum of candidate operations differentiable, and then apply the gradient descent method to directly train the weights. Inspired by DARTS \cite{liu2018darts}, Dong \emph{et al.} \cite{dong2019searching} introduce gradient-based search using the differentiable architecture sampler (GDAS) method. The authors develop a differentiable architecture sampler which samples individual architectures in a differentiable way to accelerate the architecture search procedure. Stochastic NAS (SNAS) \cite{xie2018snas} optimizes a probability distribution of the connections between different candidate operations. Li \emph{et al.} \cite{li2020sgas} observe that models with a higher performance during the search phase may perform worse in the evaluation. Hence, they divided the search process into sub-problems and proposed sequential greedy architecture search (SGAS) based on DARTS, which chooses and prunes candidate operations (e.g. convolutional layers) greedily. The authors apply SGAS for CNNs and graph convolutional networks (GCNs) and have achieved competitive performance. Xu \emph{et al.} \cite{xu2019pc} present Partially-Connected DARTS (PC-DARTS), which samples a small part of super-network to reduce the redundancy in exploring the network space. Compared with DARTS, PC-DARTS not only enjoys both faster speed and higher training stability but also a highly competitive learning performance.  Gao \emph{et al.} \cite{gao2020adversarialnas} propose the first GD-based NAS method in generative adversarial networks (GANs), called adversarial neural architecture search (AdversarialNAS), which can search the architectures of generator and discriminator simultaneously in a differentiable manner.

One bottleneck of the above GD-based NAS methods (e.g. DARTs) is that it requires excessive GPU memory during search in that all candidate network layers must be explicitly instantiated in the GPU memory. As a result, the size of the search space is constrained. To address this issue, Wan \emph{et al.} \cite{wan2020fbnetv2} propose DMaskingNAS, a memory and computationally efficient DARTS variant. DMaskingNAS employs a masking mechanism for feature map reuse. Hence, although the search space of DMaskingNAS is expanded up to $10^{14} \times$ over conventional DARTS, memory and computational costs stay nearly constant.

Another way to address the above problem is to utilize proxy tasks, e.g., learning with only a small number of building blocks or training for a small number of epochs \cite{liu2018darts,xie2018snas}. However, these approaches cannot guarantee to be optimal on the target task due to the restricted block diversity \cite{cai2018proxylessnas}. Cai \emph{et al.} \cite{cai2018proxylessnas} proposed ProxylessNAS method, which directly designs the networks based on the target task and hardware instead of with proxy. Meanwhile, the authors used path binarization to reduced the computational cost (GPU-hours and GPU memory) of NAS to the same level of normal training. Hence, ProxylessNAS algorithm can generate network architectures on the ImageNet dateset without any proxy.

Most recently GD-based NAS methods are formulated as bilevel optimization problems, However, He \emph{et al.} \cite{he2020milenas} observe that bilevel optimization in the current methods is solved based on a heuristic. For instance, solution of the problem needs to get an approximation of the second-order methods \cite{dong2019searching,liu2018darts}. He \emph{et al.} \cite{he2020milenas} demonstrate that the approximation has a superposition influence mainly because it is based on a one-step approximation of the network weights. As a result, gradient errors may cause the algorithm to fail to converge to a (locally) optimal solution. Hence, the authors propose mixed-level reformulation NAS (MiLeNAS) that uses a first-order method on the mixed-level formulation. Experimental results show that MiLeNAS has achieved higher classification accuracies than those achieved by the original bilevel optimization methods.

\section{Federated Neural Architecture Search}
Federated NAS aims to optimize the architecture of neural network models in the federated learning environment. As discussed in Section 2, distributed model training is intrinsically more difficult than centralized training, and it becomes even more challenging for NAS problems. In this section, we would like to introduce the current research on federated NAS and discuss them from two perspectives: online and offline optimization, and single- and multi-objective optimization. It should be notice that research on federated NAS work is presently limited to horizontal federated learning and federated NAS in vertical federated learning has not been reported so far.

\subsection{Offline and Online Federated Neural Architecture Search}
Most NAS methods include two steps, i.e., search the architecture of the neural network model, and training the weights of the found neural network model afterwards. And most importantly, only the final performance matters. We define these approaches as offline NAS, because the search and training steps are typically separate and only an optimized network will be used. By contrast, online NAS requires that the architecture optimization and weight training training be done at the same time, and some of the models must be used during the search process. As a result, the performance of the models during the optimization must be acceptable.

This concept can be easily extended to federated learning. In other words, federated NAS systems in which neural architecture search and weight training of the global model must be performed simultaneously are called online or real-time federated NAS, whilst federated NAS in which neural architecture search can be conducted at first and then the weights of the found models are trained are offline. Similarly, online federated NAS requires that the neural network models can be used during the optimization process.

For example, the method proposed in \cite{zhu2019multi} is a typical offline federated NAS framework using a multi-objective evolutionary algorithm. An offline evolutionary federated NAS algorithm can be summarized as follows:
\begin{enumerate}
\item Initialize parents with a population size $N$ and each individual represents one architecture of the neural network. Construct and train $N$ neural network models in federated learning with all participating clients to achieve the fitness values (e.g, validation accuracy) of the parents.
\item Generate $N$ offspring individuals by applying genetic operators on the parents. Construct and train all the generated offspring models for fitness evaluations in federated learning.
\item Combine the parent and offspring populations into one population and perform environmental selection. Select the best $N$ individual from the combined population as the new parents.
\item Repeat the above two steps until the evolutionary algorithm converges.
\item Train the weights of the optimized neural network models in federated learning.
\end{enumerate}

\begin{figure}
\includegraphics[height=7cm, width=11cm]{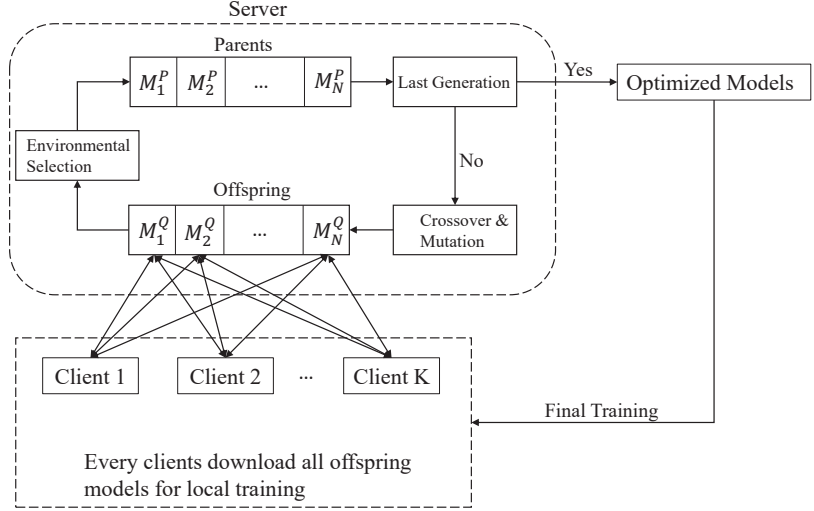}
\centering
\caption{Overall framework of offline federated NAS.}
\label{offlinefnas}
\end{figure}

It can be seen that all participating clients are used for federated model training, i.e., at each generation, all participating clients must train each of the $N$ individuals for certain rounds for fitness evaluations, which significantly increases both computation and communication costs. Client sampling can be used to alleviate this issue, in which only subsets of participating clients contribute to one individual's model training. For example in \cite{xu2020neural}, all the connected clients are divided into different groups and each sampled model use one group clients for local training. The overall process of this approach is summarized as follows:
\begin{enumerate}
\item Initialize the global model and a list of resource budgets in the server.
\item Generate a list of simplified global models by model pruning \cite{luo2017thinet} based on the current global model. And then these global models are distributed to different group clients.
\item For each communication round, every group of clients train their allocated group models for a number of pre-defined epochs and calculate the test accuracies on the validation datasets. Then both local test accuracies and validation data sizes are uploaded to the server. The server aggregates the uploaded local models and calculates a weighted accuracy for each group model. Remove $\alpha\%$ of global models with the worst test accuracies (removed global models that will not be trained and updated from the next communication round). The remaining groups of clients upload their calculated model gradients to the server for aggregation.
\item Repeat the above step for a number of pre-defined communication rounds.
\item Replace and store the global model with the first model in the global model list.
\item Repeat the above four steps until convergence.
\item Perform federated training on any stored global model.
\end{enumerate}
Although the above procedure uses different architecture generation methods (model pruning) and search space compared to the evolutionary approach, it is clearly a population based offline federated NAS framework (weight training and architecture search are separate). In addition to client sampling, the authors also remove subsets of global models to further reduce the communication costs. However, the test accuracies of each global model in the list is calculated before model aggregation, which sometimes cannot represent the real test accuracies, especially for the cases when the clients' data are particularly non-IID.

The overall framework of Offline federated NAS is shown in Fig. \ref{offlinefnas} and it has two main difficulties: 1) The number of communication rounds for federated model training of each individual is hard to determine. Setting a small number of communication rounds may make the individual's model under-trained and bias the fitness evaluations. On the other hand, setting a very large number of communication rounds consumes too many communication resources. 2) Training the candidate neural network models consume additional communication resources, which should be avoided in federated learning. For the above reasons, online federated NAS frameworks need to be developed to solve above issues.

\begin{figure}
\includegraphics[height=7cm, width=8cm]{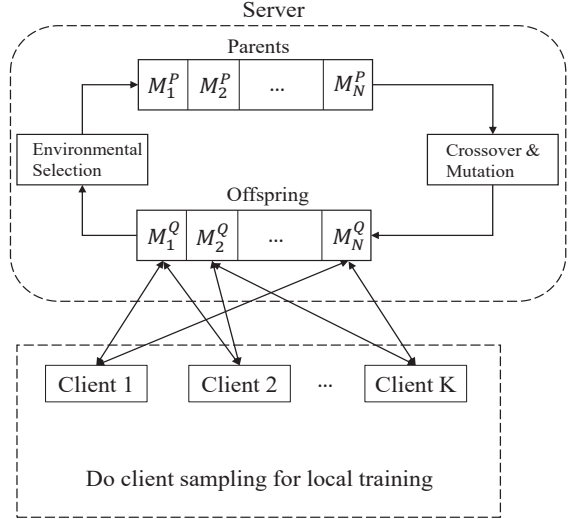}
\centering
\caption{Overall framework of online federated NAS. Client sampling are often used to ensure all offspring models are evaluated within one communication round.}
\label{onlinefnas}
\end{figure}

Online federated NAS trains the model and does the architecture optimization simultaneously (shown in Fig. \ref{onlinefnas}). To the best of our knowledge, there are currently two approaches to online federated NAS. One is gradient-based method proposed in \cite{he2020fednas}, and the other is an EA-based method proposed in \cite{zhu2020real}. The gradient based method adopts the idea of DARTS \cite{liu2018darts}, which is implemented in the federated environment. The global model here is called supernet which consists of repeated directed acyclic graphs (DAGs) and each DAG contains all candidate operations. And relaxation tricks \cite{trick1992linear} are used to make a weighted sum of the candidate operations differentiable so that the architecture parameters can be directly updated by the gradient descent algorithm. A brief description of this method is given below.
\begin{enumerate}
\item The server initializes the supernet and its architecture parameters.
\item All connected clients download the supernet and its architecture parameters from the server.
\item Each client trains and updates the supernet with fixed architecture parameters on mini-batch training data at first. Then update the architecture parameters with fixed model parameters on mini-batch validation data. These two procedures are performed alternately within one local training epoch.
\item After local training for several epochs, all participating clients upload both model and architecture parameters to the server. The server performs weighted averaging upon the supernet and architect parameters.
\item Repeat from step 2 to step 4 until convergence.
\end{enumerate}

Unlike aforementioned two offline federated NAS framework, this scheme is not population based since all candidate operations are jointly optimized. In addition, architecture search and weight training of the supernet model are conducted alternately during the period of federated training. Therefore, no additional communication resources are required for training the candidate models. However, jointly optimizing the supernet on local clients requires much more computation and memory resources, which is not well suited for the edge devices like mobile phones.

In order to reduce the memory usage of local devices, a more light-weighted real-time evolutionary NAS framework (RT-FedEvoNAS) is proposed in \cite{zhu2020real}. Different from the previous gradient based approach, RT-FedEvoNAS adopts model sampling technique \cite{guo2019single,bender2018understanding} in federated learning, where only one path of repeated cells in the global model is sampled and downloaded to local clients. As a result, both local computation and uploading costs are significantly reduced. The overall process is described as follows:
\begin{enumerate}
\item Initialize the supernet in the server. Generate the parent population containing $N$ individuals, each representing a one-path subnet sampled from the supernet using a choice key. Do client sampling to allocate $L$ clients evenly into $N$ groups.
\item Download the subnet of each parent individual to each group of clients for training. Once the training is completed, upload the $L$ local subnets to the server for aggregation to update the supernet model.
\item Generate $N$ offspring individuals using crossover and mutation. Similarly, generate a choice key for each offspring individual to sample a one-path subnet from the supernet. And then use client sampling technique to download sampled subnets (download the choice keys from the second generation) for training and uploading the trained subnets to the server for aggregation.
\item Download the supernet together with the choice keys of all parent and offspring individuals to all participating clients to evaluate the objectives. Upload all the objective values to the server and calculate the weighted average of the validation errors for each individual.
\item Combine the parent and offspring individuals into a whole population. Perform environmental selection to select $N$ new parents.
\item Repeat Step 3 to Step 4 until the generation number reaches the pre-define maximum value.
\end{enumerate}
Since only one-path of the supernet needs to be trained on each client, this sampling approach can significantly reduce both upload and local computation costs. There is a small detail in Step 4 that downloads all the supernet model to every client to calculate the validation accuracies, thus, only choice keys are downloaded in the next generation since the whole supernet has been already downloaded in the last generation.

Online methods enable federated NAS systems to perform architecture search and train the model simultaneously. Both fitness evaluation thresholds, e.g. the number of communication rounds in federated learning and extra communication resources for training the searched models are not required by using online approaches, which are highly desired for federated learning. However, the search space of online federated NAS is fairly limited, which affects the diversity of the architecture search.

\subsection{Single- and Multi-Objective Search}
The aforementioned gradient based federated NAS framework only considers and optimizes the model performances, which is usually not enough for federated learning, because federated NAS is naturally a multi-objective optimization problem. In addition to the maximization of the model performance, the payload (communication costs) to be transferred between the server and clients should be minimized. Single-objective optimization often aggregates conflict objectives into one objective using hyperparameters, while Pareto approaches aim to obtain a set of models presenting trade-off relationships between the conflicting objectives.

For example in RT-FedEvoNAS, the validation accuracy, model size and model floating point operations per second (FLOPs) of the sampled subnets are considered as the objectives need to be optimized and NSGA-II \cite{deb2002fast} is used as the basic search algorithm. Finally, after several generations of evolutionary optimization, multiple well trained subnets can be obtained chosen from the trade-off solutions based on the user's preferences.

\section{Open Challenges}
Currently, research on federated NAS is still very preliminary and several challenges remain to be solved.

\subsection{Horizontally Partitioned Data}
There is no general solution that can well solve the non-IID learning degradation problems in horizontal federated learning, let alone in federated NAS. The earliest work to explore non-IID data effect is proposed in \cite{zhao2018federated}. The authors analyze the possible reasons for divergence in global model training on non-IID data and propose a strategy to mitigate this influence by globally sharing a small part of the data across all connected clients. However, this kind of data sharing intrinsically violate the scope of privacy preserving scheme.

The federated distillation approaches \cite{jeong2018communication,lin2020ensemble} also have the potential risk of local data leakage. For the distillation, the teacher models are evaluated on mini-batches of unlabeled data on the server and their logits for mini-batch are used to train the student model on the server. The server can get a lot of local data information even on fake mini-batch data generated by local GAN generators \cite{goodfellow2014generative}.

Some statistical aggregation methods \cite{mohri2019agnostic,li2018federated} are proposed to replace the original federated averaging algorithm (FedAvg). Both mathematical and experimental results prove that the proposed aggregation algorithms outperform the FedAvg on non-IID data. However, these approaches are often limited to some specific models and datasets, and it is unclear if they can show better performance for federated NAS frameworks. Hsieh \emph{et al.} discuss the effect of non-IID data for DNNs in detail and different federated optimization methods are used upon different DNNs, such as GoogleNet \cite{szegedy2015going},and ResNet \cite{he2016deep}. Experiment results show that batch normalization \cite{ioffe2015batch} performs really poorly on non-IID data, but batch renormalization \cite{ioffe2017batch} and group normalization \cite{wu2018group} are much more robust for non-IID data, which are much more suited for federated learning. Most recently, it is shown in \cite{xu2020} that ternary quantization is helpful in alleviating model divergence in federated learning, although its effectiveness remains to be validated on federated NAS.

\subsection{Vertically Partitioned Data}
Current federated NAS methods are all based on horizontal federated learning. Unlike horizontal federated learning, it is really hard to determine whether the data are IID or non-IID in vertical federated learning, since they are 'partitioned' towards the feature space.

Most existing vertical federated learning frameworks are built on two-party systems using simple linear models. Since only one party can hold the labels, the loss needs to be calculated on ciphertext; otherwise, the label information will be revealed. Then the gradients are very hard to calculate since the total loss is encrypted. Some approximation techniques like Taylor expansions \cite{hardy2017private,liu2020secure} are often used to simplify the gradient calculations, which, however, may introduce strong biases for complex models like DNNs.

Overall, vertical federated NAS is totally different from horizontal federated NAS, which is in general still an unexplored research area.

\subsection{Adversarial Federated Neural Architecture Search}
Adversarial federated learning has two purposes: 1) Inference of the client data information 2) Attack the global model to conduct backdoor \cite{pmlr-v108-bagdasaryan20a} elements or even let the model unusable. And the adversary in federated learning can be one of participating clients or the central server, since we often assume the server is honest-but-curious. Thus, the server should also be considered as a potential risk.

Federated learning is still fragile to white box attacks, since the model gradients and parameters still contain local data information. Geiping \emph{et al.} \cite{geiping2020inverting} showed that local images can be reconstructed from the knowledge of model parameters (or gradients) by inverting gradients techniques. In addition, an adversarial GAN \cite{goodfellow2014generative} generator can be developed on either the server \cite{wang2019beyond} or the client side \cite{hitaj2017deep}. The adversary can reconstruct other participating clients' private data, even it has no knowledge of the label information.

Enthoven and Al-Ars \cite{enthoven2020overview} summarize most defence strategies used in federated learning, which can be categorized into three types: 1) Subsample or compress the communicated gradients \cite{shokri2015privacy,konevcny2016federated,yoon2020federated}; 2) Differential privacy and SMC \cite{cryptoeprint:2017:281}, and 3) Robustness aggregation \cite{pillutla2019robust} using e.g., the byzantine resilient aggregation rule \cite{lamport2019byzantine,mhamdi2018hidden}.

In general, finding robust model architectures in federated learning to defend against the adversarial attacks is still a hard task.

\subsection{Encrypted Federated Neural Architecture Search}
Homomorphic encryption technologies are often applied to prevent privacy leakage from the gradient information sent to the server. However, using homomorphic encryption in federated NAS system has two main difficulties.

First, homomorphic encryption, including encoding and encryption, is computationally expensive in federated learning.  At first, all communicated model parameters need to be encoded into large inter numbers, because homomorphic encryption does not work on real numbers. Then the encoded parameters need to do modulus calculations with large prime numbers one by one. Unfortunately, modern deep neural network models contain millions of parameters, making the encryption process computationally extremely intensive. Therefore, developing a light weighted encryption method is an important yet challenging task for federated learning, let alone for federated NAS.

Second, the original federated encryption is introduced in which the server holds the public key and clients hold the private key. This framework is unsafe, because only one of clients uploads its secret key to the server. Therefore, a more advanced SMC approach is adopted that divides the whole secret key into several shards, and the server cannot decrypt the gradients unless it collected $t$ (secret key recover threshold) key shards.  Unfortunately, the encrypted gradients must be frequently transferred between the server and clients, which significantly increases the communication costs, since the local clients can only partially decrypt the gradients through their key shards. This is a big burden to the communication resources, which needs to be solved in the future.

\section{Conclusion}
In this survey paper, a brief overview of federated learning and NAS is provided, and a combination of both techniques, i.e., federated NAS is introduced. Given several remaining challenges in both federated learning and NAS, federated NAS becomes extremely challenging since many techniques developed in centralized NAS are no longer suited for federated NAS, and NAS will be subject to more constraints introduced by the federated learning environment. Two approaches to federated NAS are discussed, one is offline optimization and the other is online optimization. It is noted that offline evolutionary NAS methods are not applicable for many real-world scenarios, mainly because the offline approach performs architecture search and weight training separately and requires a large amount of communication costs. In addition, the performance of neural network under optimization must be acceptable for application and serious performance drop is not allowed. RT-FedEvoNAS \cite{zhu2020real} offers a solution to the above challenges, although its search space is highly constrained.

Despite that many grand challenges remain to be solved, federated NAS is of paramount practical significance for many real-world problems, where handcrafted deep neural networks may fail to work properly. We hope that this survey will help understand the promises and challenges in federated NAS, thereby triggering more research interests in developing new theories and algorithms, thereby promoting the application of AI techniques to a wider range of fields where data privacy and security is a primary concern.

\bibliographystyle{unsrt}
\bibliography{reference}

\end{document}